\documentclass[aps,prl,reprint,groupedaddress]{revtex4-1}

\usepackage{graphicx}
\usepackage[margin=1in]{geometry}
\usepackage{amsmath}

\begin{document}

\title{Suppression of Three-Body Loss Near a \textit{p}-Wave Resonance \\ Due to Quasi-1D Confinement}

\author{Andrew S. Marcum}
\author{Francisco R. Fonta}
\author{Arif Mawardi Ismail}
\author{Kenneth M. O'Hara}
\affiliation{Physics Department, The Pennsylvania State University, 104 Davey Lab, University Park, PA 16802, USA}

\date{\today}

\begin{abstract}
We investigate the three-body recombination rate of a Fermi gas of $^6$Li atoms confined in quasi-1D near a $p$-wave Feshbach resonance. We confirm that the quasi-1D loss rate constant $K_3$ follows the predicted threshold scaling law that $K_3$ is energy independent on resonance, and find consistency with the scaling law $K_3 \propto (k \, a_{1D})^6$ far from resonance [Mehta {\emph{et al.}} Phys. Rev. A {\bf{76}}, 022711 (2007)]. Further we develop a theory based on Breit-Wigner analysis that describes the loss feature for intermediate fields. Lastly we measure how the loss rate constant scales with transverse confinement and find that $K_3 \propto V_L^{-1}$, where $V_L$ is the lattice depth. Importantly, at our attainable transverse confinements and temperatures, we see a 74-fold suppression of the on-resonant three-body loss rate constant in quasi-1D compared to 3D. With significant further enhancement of the transverse confinement, this suppression may pave the way for realizing stable $p$-wave superfluids.
\vspace{-1.0in}
\end{abstract}

\maketitle

\vspace{-0.5in}
\vspace{-0.25in}$S$-wave Feshbach resonances have been used with great success to study the smooth BEC to BCS crossover in dilute Fermi gases.  A promising approach to realize {\emph{unconventional}} superfluidity in dilute ultracold atomic gases is to investigate pairing in a spin-polarized Fermi gas near a $p$-wave Feshbach resonance~\cite{Ticknor2004,Schunck2005,Chevy2005}.  In contrast to conventional BCS superfluids with condensates comprised of spin-singlet Cooper pairs with isotropic ($s$-wave) pair wavefunctions, unconventional superfluids feature non-trivial anisotropic pairing with correspondingly rich phase diagrams and exotic quasiparticle excitations~\cite{GurarieResPairedSF}.  In 3D, such a system is predicted to exhibit an array of phases separated by classical, quantum, and topological phase transitions~\cite{GurarieResPairedSF,Gurarie2005,Botelho2005,Cheng2005, Iskin2006}.  In reduced dimensions, $p$-wave superfluids have remarkable properties of current intense interest.  For example, in 2D a topological $p_x + i p_y$ superfluid characterized by a Pfaffian ground state with non-Abelian excitations is expected \cite{GurarieResPairedSF,Read2000,Fedorov2017,Tewari2007}.  In 1D, a spin-polarized Fermi gas with $p$-wave pairing may provide a realization of Kitaev's chain which can feature unpaired Majorana fermions localized at the ends of the chain~\cite{Kitaev2001}.

Unfortunately, inelastic loss near $p$-wave Feshbach resonances has compromised attempts to observe $p$-wave superfluidity in 3D.  Spin-polarized Fermi gases that are not in their lowest energy hyperfine state suffer from strong two-body dipolar relaxation.  While such loss can be avoided for fermions in their absolute ground state, three-body recombination rates are still large enough to prohibit evaporation to degeneracy at equilibrium~\cite{Regal2003,Zhang2004,Waseem2018,Deh2008}.  Only out-of-equilibrium studies of the $p$-wave contacts have been possible~\cite{Thywissen16}.

While three-body recombination (3BR) has proven insurmountable in 3D, it has been predicted to be suppressed for atoms confined to quasi-1D. First, Mehta, Esry, and Greene found the threshold scaling laws for the three-body recombination in 1D. They predicted the on resonance 3BR rate constant to be independent of energy, $K_3 \propto {\mathrm{const.}}$, and the far from resonance 3BR rate constant to scale as $K_3 \propto (k \, a_{1D})^6 $ where $k$ is the relative wavenumber and $a_{1D}$ is the 1D scattering length~\cite{1D3BR}. When contrasted with the  3BR rate constant in 3D~\cite{Suno2003,Suno_2003}, the 1D scaling laws imply a significant reduction in three body loss at low temperature ~\cite{Shlyapnikov}. More recently, Zhou \& Cui have shown that the shallow molecules induced near a $p$-wave Feshbach resonance are significantly more spatially extended in quasi-1D compared to their 3D counterparts which further suggests a suppression of 3BR in quasi-1D~\cite{Zhou2017}.

In this work we measure the 3BR loss rate in both 3D and quasi-1D for a $^6$Li gas spin-polarized in the lowest hyperfine state ($\left| 1 \right\rangle$) and in the vicinity of the $p$-wave Feshbach resonance (FR) at $159.1 \, {\mathrm{G}}$\cite{Schunck2005,Chevy2005}. At our lowest attainable temperatures, we observe a 74-fold suppression of the loss rate constant on resonance in quasi-1D relative to that in 3D. We go on to confirm the on-resonance scaling law and corroborate the off-resonance scaling, which mirrors that of even-parity bosons in quasi-1D~\cite{Zundel, 1D3BR}.  Further, we develop a theory to explain the observed loss  at intermediate fields based on Breit-Wigner analysis.  Finally, we find the 3BR loss rate constant to be inversely proportional to the depth of the 2D lattice used to constrain the atoms to quasi-1D.

To prepare low-temperature samples that can be loaded into a 2D optical lattice, we begin with $^6$Li atoms confined in a crossed optical dipole trap (CODT) formed by one 1064~nm and one 1070~nm laser beam intersecting at an angle of $12^{\circ}$.  Each beam is focused to a waist of $30\,\mu{\mathrm{m}}$ at the point of intersection and contains up to 80~W of power.  A balanced mixture of the two lowest energy hyperfine levels (labeled $|1\rangle$ and $|2\rangle$) is loaded from a gray optical molasses~\cite{Burchianti2014,Sievers2015} into the CODT at full power.  Forced evaporative cooling brings the atoms to the experimentally desired temperature.

\begin{figure}[tb]
	\centering
	\includegraphics[width=\columnwidth]{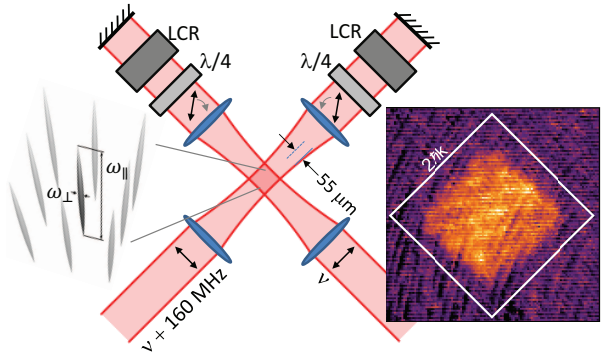}
	\caption{\label{OptLattFig} Schematic view of experimental setup as well as an example band mapping image showing all atoms in ground band of the lattice.}
    \vspace{-0.225in}
\end{figure}

This sample can be loaded into a 2D optical lattice formed by two orthogonal pairs of retroreflected laser beams, both at 1064 nm, as shown in Fig.~\ref{OptLattFig}. Each of these four lattice beams is focused to a horizontal (vertical) waist of 55 (300) $\mu$m. A liquid crystal retarder and quarter waveplate are included in each arm to act as dynamically adjustable polarization rotators for the retroreflected polarizations. This makes it possible to adiabatically switch between a 3D trapping configuration and the optical lattice configuration, a feature used to load atoms into the optical lattice\cite{Liao2010}. A frequency offset of 160 MHz eliminates interference between the beam pairs.

\footnotetext[1]{The field-dependent scattering volume $v_p$  has the resonant form $v_p =  V_{\mathrm{bg}} \Delta B/(B - B_{\mathrm{res}})$.}

\footnotetext[2]{We use  $k_e = 2 \hbar^2/(m \, V_{\mathrm{bg}} \, \Delta B \, \delta\mu)$~\cite{GurarieResPairedSF} with $V_{\mathrm{bg}} \, \Delta B = - 2.8 \times 10^6 \, a_0^3$ from Ref.~\cite{Waseem} and $\delta \mu = k_B (113 \mu{\mathrm{K}}/{\mathrm{G}})$  (the relative magnetic moment between the molecular state and the atomic state)  from Ref.~\cite{Fuchs2008}.}

\footnotetext[3]{Here $k_e$ is defined consistent with the effective range expansion $k^3 \, \cot \delta_p = -1/v_p + (1/2) \, k_e \, k^2$ of the scattering phase shift $\delta_p$.}

\footnotetext[5]{ This is consistent with less frequent measurements of $\omega_{\perp}$  by parametric resonance.}
With the lattice beams at full power, the transverse site trapping frequency is measured on a daily basis via time of flight (TOF) expansion  to be $\omega_{\perp} = 2 \pi \times 281(4)$ kHz~\cite{Note5}.   This corresponds to a lattice depth of $23 E_R$, where $E_R$ is the recoil energy, and a tunneling time of $\tau = 22$ ms.   At this depth we observe a confinement induced resonance (CIR) shift of $120(5)\,{\mathrm{mG}}$ relative to the location of the 3D resonance. This is in excellent agreement with the predicted CIR shift~\cite{ Shlyapnikov, Zhou2017} of $120\,{\mathrm{mG}}$ for a $23\,E_R$-depth lattice using the scattering parameters $V_{\mathrm{bg}} \, \Delta B = -2.8 \times 10^6 \, a_0^3$~\cite{Nakasuji2013,Note1},  and $k_e = -0.182 \, a_0^{-1}$~\cite{Waseem, GurarieResPairedSF,Fuchs2008,Note2} where $V_{\mathrm{bg}} \Delta B$ is the product of the background scattering volume and resonance width, and $k_e$ is the effective range~\cite{Note3}.  Breathing mode spectroscopy is used to measure a longitudinal site frequency of $\omega_{||} = 2 \pi \times 300(10)$Hz. For atoms loaded into the lattice, at maximum we measure $T = 4.6\; \mu$K and $T_F = 2 \mu$K. Thus in all cases the atoms in the lattice satisfy $k_b T,\; k_b T_F,\; \hbar \omega_{||}  <  \hbar \omega_{\perp}$ and each lattice site can be treated as an individual quasi-1D trap.

Establishing high magnetic field stability is a prerequisite to accurately determine the thermally averaged 3BR rate constant $L_3$ near the narrow $p$-wave resonance.   By actively stabilizing the field we achieve residual rms field fluctuations $<4 \, {\mathrm{mG}}$ as indicated by performing radio-frequency (RF) spectroscopy on the $\left| 1 \right\rangle \rightarrow \left| 2 \right\rangle$ transition.


The thermally averaged 3BR rate constant $L_3$ is first measured in a 3D trap provided by the forward propagating lattice beams only (the retro-reflected beams are blocked to guarantee the potential is not corrugated).  The atoms are loaded into this 3D configuration of the lattice beams and held for 1 s to allow them to equilibrate in the new trap.  With $\approx 5 {\mathrm{W}}$ per beam, a cigar shaped trap with trap frequencies $\omega_x = \omega_z = 2\pi \times (730\,{\mathrm{Hz}})$ and $\omega_y = 2\pi \times (180\,{\mathrm{Hz}})$ is formed.  A spin polarized gas is created by ramping the magnetic field to the location of the $|1\rangle$-$|1\rangle$ FR in 20 ms, followed by a hold time of $100 \, {\mathrm{ms}}$, during which the entire state $|1\rangle$ population decays via 3BR. After this clearing there are $4 \times 10^5$ atoms in state $|2\rangle$ at $T = 1.8(1) \; \mu\text{K}$ and $T/T_{F} = 0.6$, where $T_{F}$ is the Fermi temperature and the gas can be considered a thermal gas. The magnetic field is then ramped to the field of interest (FOI) in 1 ms, followed by a 20 ms wait time which allows the magnetic field to stabilize. The atoms are then transferred to state $|1\rangle$ using a RF pulse, held for a time $t$, and transferred back to state $|2\rangle$ with a second, identical RF pulse. The double RF pulse technique is required to avoid unwanted decay during the field ramps to and away from the FOI. The \textit{in situ} density profile of the cloud is then imaged using phase contrast imaging from which total atom number and temperature are extracted as a function of hold time $t$.

\begin{figure}[tb]
	\centering
	\includegraphics[width=\columnwidth]{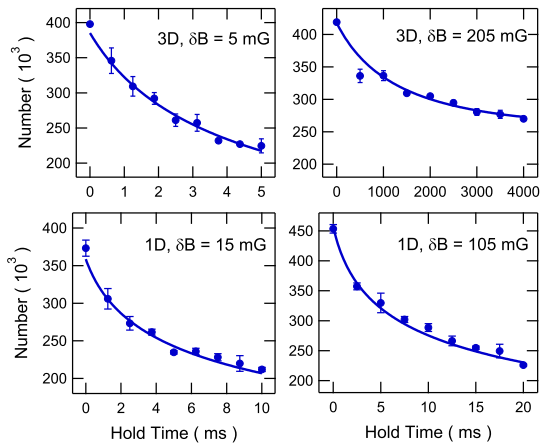}
	\caption{\label{DecayCurvesFig} Example decay curves used to extract the loss rate in 3D and quasi-1D.}
    \vspace{-0.225in}
\end{figure}

\footnotetext[4]{Data points appearing throughout are averages of 3 to 5 individual measurements; error bars are the standard error of the mean.}

The quasi-1D measurement follows a procedure very similar to what is used in 3D. The atoms are loaded into the 3D configuration of the lattice beams, held for 1 s, and state $|1\rangle$ is cleared as above. The polarizations of the lattice beams are then ramped to the optical lattice configuration, and the lattice beam power is increased to its maximum over 200 ms.  At this stage, we confirm the entire atomic population resides within the ground band of the lattice via band mapping, the result of which is also shown in Fig. \ref{OptLattFig}. The measurement of $L_3$ is completed using the same double RF pulse technique used in the 3D case.

We extract $L_3$ by fitting the atom number remaining after time $t$ to a three-body loss curve, which is given by the rate equation
\begin{equation}\label{RateEqn}
\frac{\dot{N}}{N} = - L_3 \langle n^2 \rangle,
\end{equation}
where $\langle n^2 \rangle$ is the mean squared density. Fig. \ref{DecayCurvesFig} shows example decay curves in both 3D and 1D~\cite{Note4},  for several detunings from the \textit{p}-wave FR. The 1D data cannot be fit directly with the result of Eqn. (\ref{RateEqn}), since we observe an array of many tubes, and because the initial number in a given tube varies according to the 3D density profile of the cloud. To account for this, we assume the decay in each tube is still governed by Eqn. (\ref{RateEqn}) and calculate the total number of atoms remaining at time $t$ by summing over the individual tubes.  In addition, we assume the lattice site frequencies are constant across all occupied tubes.

\begin{figure}[tb]
	\centering
	\includegraphics[width=\columnwidth]{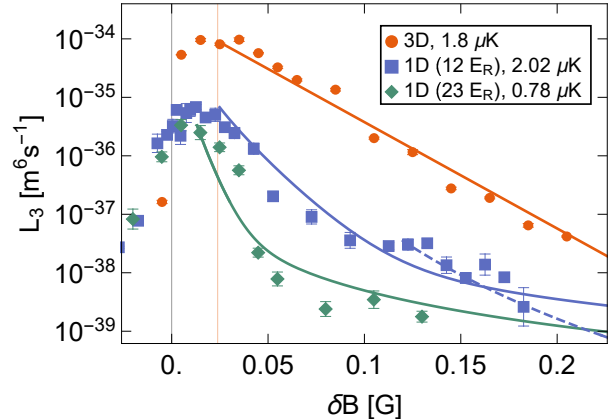}
	\caption{\label{L3vsB} Magnetic field dependence of $L_3$ in 3D and quasi-1D. The two 1D data sets were taken at different lattice depths, resulting in differing CIR shifts. For clarity, all the data sets are shifted so that the resonance locations overlap at $0 \, \delta B$ . The colored vertical line shows the field below which $L_3$ is expected to be unitarity limited in 3D. The solid red curve is the intermediate theory of Ref.~\cite{Waseem2019} fit to our 3D data. The solid blue and green curves are Eqn.~\ref{Eqn:IntmdL3Theory} fit to our quasi-1D data sets. The dashed curve shows the far from resonance 1D scaling law for comparison.  Most notably, the resonant loss rate constant in the deep quasi-1D trap is reduced by a factor of 29 relative to that in 3D.}
\vspace{-0.225in}
\end{figure}

As shown in Fig.~\ref{L3vsB}, in 3D we see good agreement between the dependence of $L_3$ on field detuning and the theoretical predictions.  Far from resonance, beyond the range of Fig.~\ref{L3vsB}, $L_3$ has been shown to scale as $v_p^{8/3}$ in agreement with the expected threshold scattering behavior \cite{Suno2003}. Nearer to resonance, $L_3$ is described by an intermediate theory (solid red curve in Fig.~\ref{L3vsB}) based on rate equations developed and detailed by Waseem \textit{et al.} \cite{Waseem2019}. In this region
\begin{equation}
L_3 \approx 9 K_{AD} (6 \pi/k_T^2)^{3/2}e^{-k_{\mathrm{res}}^2/k_T^2} ,
\end{equation}
where $K_{AD}$ is the atom dimer relaxation coefficient, $k_T = (3 m k_B T/2 \hbar^2)^{1/2}$ is the thermal wavenumber and $k_{\mathrm{res}} = (|v_p|k_e)^{-1/2}$ is the wavenumber for resonant scattering in the continuum for a given scattering volume $v_p$. From our intermediate regime 3D data we obtain $K_{AD} = 6.5(1.0) \times 10^{-17} \text{m}^3/\text{s} $. We attribute the difference between our measurement and that obtained by Waseem \textit{et al.} to the different temperatures at which the measurements were made. The coefficient $K_{AD}$ has already been shown to have a temperature dependence for $s$-wave resonances \cite{Li2018} suggesting that similar behavior for $p$-wave resonances may be expected.

Very near resonance, $L_3$  becomes unitarity limited and independent of the detuning from resonance \cite{Suno2003}. The unitary regime was seen to occur in Ref.~\cite{Waseem2019} for $k_T/k_{\mathrm{res}} \ge 1$. In this region, $L_3$ is limited to its maximum value.  We observe similar behavior in Fig.~\ref{L3vsB}, consistent with the unitarity limit.

Fig. \ref{L3vsB} also shows the field dependence of $L_3$ in 1D at $T = 0.76(3)$ and $2.02(3) \; \mu$K. Mehta, Esry, and Greene predict that far from resonance $L_3$ should scale as $a_{\mathrm{1D}}^6$ \cite{1D3BR} and, for comparison to our $2.02 \mu{\mathrm{K}}$ data, we plot this prediction (dashed blue line). Unfortunately, residual heating in the lattice prohibits us from employing the necessary observation times to definitively confirm this scaling law dependence by measuring $L_3$ even further from resonance.  For intermediate fields we find good agreement with a Breit-Wigner theory developed below.  Most crucially, we observe that the on-resonant value of $L_3$ in 1D is suppressed by up to a factor of 29 (for the $V_L = 23 \, E_R, 2.02 \mu{\mathrm{K}}$ data) as compared to the on-resonant value in 3D.  We explore the temperature and lattice depth dependence of this on-resonant suppression below.

Here we first develop an intermediate theory to explain our quasi-1D 3BR  loss feature for intermediate fields based on Breit-Wigner scattering theory \cite{TaylorBook,Waseem2018,Hazlett12,Napolitano94,Mathey09,Miles00,Yurovky03}. This theory assumes that the dominant three body loss mechanism is derived from two particles resonantly forming a quasi-bound molecule which subsequently decays to a deeper molecular state upon collision with a third atom. The inelastic cross section takes the Briet-Wigenr form
\begin{equation}
 \sigma_{p_{1d}}^{in} =\frac{3 \pi}{k^{2}} \frac{\Gamma_{e_{1D}}\Gamma_0}{(E-E_{\mathrm{res}})^2 +\frac{(\Gamma_{e_{1D}}+\Gamma_0)^2}{4}},
\end{equation}
where $E_{\mathrm{res}}$ is the  energy of the quasi-bound molecule, $\Gamma_{e_{1D}}$ is its resonant energy width, and $\Gamma_0/\hbar$ is the inelastic atom-dimer relaxation rate. We can express the inelastic energy width as $ \Gamma_0 = \hbar K_{AD} n $ where $K_{AD}$ is the atom-dimer relaxation coefficient, and $n$ is the density. The atom loss can then be expressed as
\begin{equation}
	\dot{n} = -\frac{3}{6}\frac{2 \hbar k}{m} \sigma_{p_{1d}}^{in} n^2 = -K_3 n^3,
\label{Eqn:nDotLossRate}
\end{equation}
where
\begin{equation}
	K_3 = 3\frac{\pi \hbar}{m k} \frac{\Gamma_{e_{1D}}K_{AD}}{(E-E_{\mathrm{res}})^2 +\frac{(\Gamma_{e_{1D}})^2}{4}}.
\label{Eqn:K3}
\end{equation}
A factor of $3/6$ in Eqn.~\ref{Eqn:nDotLossRate} has been added since every inelastic collision event results in 3 lost atoms and there are $N^3/6$ triplets per unit volume. In Eqn.~\ref{Eqn:K3} we assume that we are sufficiently detuned from resonance such that $\Gamma_0 \ll \Gamma_{e_{1D}}$.

\footnotetext[6]{The quasi-1D $S$-matrix for $p$-wave scattering is given by $S_{1D} = \left(-\frac{1}{a_{1D}}+ \frac{1}{2}r_{1D} k^{2} + i k \right)/\left(-\frac{1}{a_{1D}}+ \frac{1}{2}r_{1D} k^{2} - i k \right)$.}

While this mirrors the development of a 3D theory \cite{Waseem2018} all parameters are derived from the quasi-1D $S$-matrix as opposed to the 3D $S$-matrix. For quasi-1D, $a_{1D}$ and $r_{1D}$ (the 1D effective range) replace $v_p$ and $k_e $ in the 3D $S$-matrix.  Further, the quasi-1D $S$-matrix for $p$-wave scattering has the same $k$ dependence as the $S$-matrix for 3D $s$-wave scattering~\cite{Note6}.  Thus, instead of the usual $p$-wave resonant energy width $\Gamma_{e_{3D}} \propto E^{3/2} $, we have  $\Gamma_{e_{1D}} = \sqrt{4 \hbar^2 E /(m r_{1D}^2) } $ as in the $s$-wave case. Similarly, the quasi-bound state energy is derived from the $S$-matrix to be $E_{\mathrm{res}} = 2 \hbar^2 / (m a_{1D} r_{1D}) $.

Finally, to obtain $L_3$, we take the thermal average of $K_3$  in one dimension,
\begin{equation}
	L_3 = \frac{1}{\sqrt{\pi k_B T}}  \int_{0}^{\infty} \frac{K_3}{\sqrt{E}} e^{-\frac{E}{k_B T}} dE.
\end{equation}
By taking the limit $\Gamma_{e_{1D}}  \ll k_B T$ (which holds true for all our measured temperatures), and performing integration by parts to separate the contributions due to the resonance embedded in the continuum and the diverging density of states in 1D at low energy, we find an analytic solution to the integral:
\begin{equation}
 L_3 = \frac{3 \pi \hbar^3}{m^{3/2}} K_{AD}\left[\frac{\sqrt{\frac{4 \hbar^2} { m r_{1D}^2} }}{E_{res}^2}+ \frac{2 \pi e^{-\frac{E_{res}}{k_B T}}}{E_{res}\sqrt{\pi k_B T}} \right] .
\label{Eqn:IntmdL3Theory}
\end{equation}
In this form we can identify two components to the three body loss; the first term is the low energy contribution and dominates far from resonance, the second term is derived from the resonance embedded in the continuum and dominates nearer to resonance.  Very near resonance in quasi-1D $L_3$ becomes unitarity limited as we observe that $L_3$ becomes constrained to a maximum value near resonance.  It is only beyond the unitary regime that we expect our intermediate theory to be accurate.

The solid lines fitting the 1D data in Fig. \ref{L3vsB} are single parameter fits to Eqn.~\ref{Eqn:IntmdL3Theory} with $K_{AD}$ as the only unknown. Only data outside an empirically determined unitarity-limited region are included in the fit (indicated by horizontal limits of the solid lines).  The fits yield a value of $K_{AD}= 14(2) \times 10^{-17} {\mathrm{m}}^3/{\mathrm{s}} $  at 2.02 $\mu$K, and  $K_{AD}= 3.6(1.6) \times 10^{-17} {\mathrm{m}}^3/{\mathrm{s}} $ at 0.78 $\mu$K.  This intermediate theory provides a good description of the data outside the unitarity-limited region and before the far-off-resonance scaling limit.

\begin{figure}[tb]
  	\centering
   	\includegraphics[width=\columnwidth]{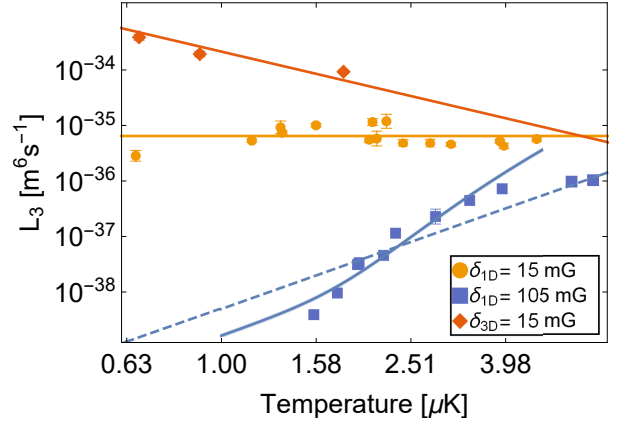}
    \caption{\label{L3VsT} Log-log plot of $L_3$ versus temperature. Solid red curve shows 3D unitary limit. Solid orange line shows the on resonance $L_3 \propto {\mathrm{const.}}$  scaling law. Dashed blue curve shows the $L_3 \propto T^{3}$ scaling law. Solid blue curve is Eqn.~\ref{Eqn:IntmdL3Theory} with no free parameters assuming $K_{AD} \propto T^3$.}
    \vspace{-0.225in}
\end{figure}

Next we sought  to confirm the energy dependence of the scaling law predictions of Ref.~\cite{1D3BR} i.e. in 1D, $L_3$ scales as $E^3$ far from resonance and is energy independent  very near resonance.  To study this we measure the temperature dependence of $L_3$ in 1D at two detunings from the 1D resonance position: $\delta_{\text{1D}} = 15$ mG and $105$ mG. By varying the endpoint of evaporative cooling, we achieve temperatures of the atoms in 1D between $0.66$ and $4.6\; \mu{\mathrm{K}}$.

Fig.~\ref{L3VsT} shows our measured energy dependence of $L_3$ in the $V_L = 23 E_R$ lattice.  Fitting the on-resonance values of $L_3$ to a temperature-dependent power law $L_3 \propto T^p$ yields $p = -0.04(5)$, consistent with $L_3 = {\mathrm{const.}}$, confirming the on-resonance threshold scaling law of Ref.~\cite{1D3BR}.  We find the on-resonant constant value $L_3 = 6.5(2) \times 10^{-36} {\mathrm{m}}^6/{\mathrm{s}}$ (orange solid line) at this lattice depth.    This energy independence is in stark contrast to the on-resonance 3D loss rate that has been shown~\cite{Waseem2018} to have a unitarity limited scaling $L_3 \propto T^{-2}$ which we observe again here with $L_3 = \left(2.1(1) \times 10^{-46} {\mathrm{K}}^2 {\mathrm{m}}^6 /{\mathrm{s}} \right) \times T^{-2}$ (red solid line).     Based on the on-resonance fits in quasi-1D and 3D and for the lowest temperature we attain in quasi-1D (660 nK), we find the on-resonant value of $L_3$ is suppressed by a factor of 74(4) relative to that in 3D.  Finally, we fit the off-resonance quasi-1D data to a $T^3$ dependence (dashed line) which captures the trend in the data but also shows significant deviations.  As we only expect a pure $T^3$ dependence in the far-off-resonance limit, we find better agreement if we instead fit the data with Eqn.~\ref{Eqn:IntmdL3Theory} (solid line) where a $K_{AD} \propto T^3$ scaling is included for consistency with the expected far-from-resonance threshold behavior ($L_3 \propto T^3$) as described below.

To fit our Breit-Wigner theory to the temperature dependent data in Fig.~\ref{L3VsT} we need to know the temperature dependence of $K_{AD}$.   Since the first term in Eqn.~\ref{Eqn:IntmdL3Theory} (which dominates far from resonance) has no explicit temperature dependence, all of the far from resonance temperature dependence  must come directly from $K_{AD}$ and we thus assume $K_{AD} \propto T^3$. This is consistent with all values of $K_{AD}$ measured to date. Indeed, if we fit $K_{AD} \propto T^p$ for the values  determined from the quasi-1D and 3D data in Fig.~\ref{L3vsB} as well as 3D measurements reported by Waseem \textit{et al.}~\cite{Waseem2019} at higher temperature, we find $p = 3.55(46)$, consistent with a $T^3$ dependence.  Further, if we fit this same data to a fixed $T^3$ scaling to determine the multiplicative pre-factor, we find $K_{AD} = \left( 1.4(2) \times 10^{-17} {\mathrm{m}}^3/{\mathrm{s K}}^3\right) \times T^3$.  Using this result for $K_{AD}$  allows us to plot Eqn.~\ref{Eqn:IntmdL3Theory} with no free parameters (solid blue line) in Fig.~\ref{L3VsT}.  The agreement is good up until the highest temperature data where the theory must break down as it approaches the unitarity limit.

\begin{figure}[tb]
	\centering
	\includegraphics[width=\columnwidth]{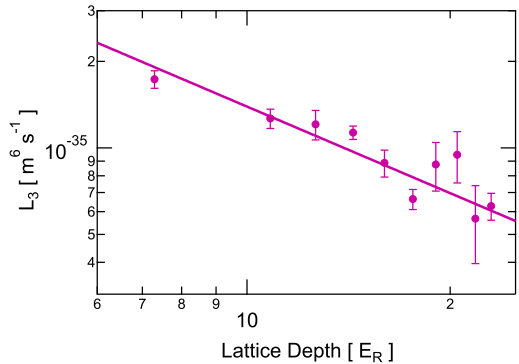}
	\caption{\label{L3VsER}Log-log plot of on resonance $L_3$ vs lattice depth. Solid line shows $L_3 \propto V_L^{-1}$ scaling.}
    \vspace{-0.225in}
\end{figure}

Finally, having demonstrated that $L_3$ is independent of energy on resonance in agreement with theory, we determined the dependence of the on-resonant value of $L_3$ on confinement strength. Fig.~\ref{L3VsER} shows the dependence of $L_3$ on lattice depth $V_L$ obtained when the magnetic field detuning was kept within $\delta B < 15\,{\mathrm{mG}}$ of resonance.  This data was fit to a power law $L_3 \propto V_L^p$ where $p = -0.92(8)$.  Thus, our data is consistent with an on-resonant value of $L_3$ that varies inversely with lattice depth and, correspondingly, in proportion to $a_\perp^4$ (where $a_\perp$ is the transverse harmonic oscillator length).   Thus, significant suppression should be attainable with increased  transverse confinement.

In conclusion we have measured the 3BR rate constant in quasi-1D and shown its scaling with field, temperature, and transverse confinement. Importantly, the on resonance 3BR rate constant was shown to be independent of temperature, resulting in very strong suppression relative to its 3D counterpart at low temperatures. The on resonance loss rate constant was further shown to scale with $a_\perp^4$ suggesting that considerable suppression of the loss rate constant should be possible by substantially increasing the transverse confinement. To this end, as a promising direction for future research, we expect  that the use of a very deep 2D square lattice made from retro-reflected 532 nm light will result in significant suppression of three body loss and may even permit the stabilization of a $p$-wave superfluid.

{\emph{Note added.}} -- During manuscript preparation we became aware that 3BR near a $p$-wave FR in quasi-1D has also been investigated by the Rice group~\cite{Hulet2020}.  Their conclusion that 3BR is not suppressed in quasi-1D relies on reporting the loss rate constant relative to the 1D density.  However, the 3D density is the relevant density as scattering is still three-dimensional even for the tightest confinement, i.e. $a_\perp \gg \ell_{\mathrm{vdW}}$ or equivalently $E_{\mathrm{vdW}} \gg \hbar \omega_\perp$ where $\ell_{\mathrm{vdW}}$ ($E_{\mathrm{vdW}}$) is the van der Waals length (energy).

\end{document}